\documentclass[twocolumn]{article}
\RequirePackage{xcolor}
\usepackage[a4paper, total={18cm, 26cm}]{geometry}
\usepackage{amsmath,amssymb,amsfonts}
\usepackage{algorithmic}
\usepackage{algorithm}
\usepackage[outline]{contour}
\usepackage{hyperref}
\usepackage{graphicx}
\usepackage{siunitx}
\usepackage{pgfplots}
\pgfplotsset{compat=newest}
\usepackage{tikz}
\usetikzlibrary{spy}
\usepackage{tikz-3dplot}
\usepackage{circuitikz}

\usetikzlibrary{shapes,arrows}
\usepackage{microtype}

\DeclareMathOperator{\spanS}{span}

\begin{document}

\title{\textbf{A Spline-based Partial Element Equivalent Circuit Method for Electrostatics}}

\author{Riccardo Torchio, Maximilian Nolte, Sebastian Schöps, Albert E. Ruehli\thanks{This work is supported by the Graduate School CE within the Centre for Computational Engineering at TU Darmstadt, by the German Research Foundation via the project 443179833 and the DAAD in the framework of Short-Term Grants (57588366). The authors thank Felix Wolf, Jürgen Dölz, and Stefan Kurz for the fruitful discussions on the topic. ({\it Corresponding author}: Riccardo Torchio). Riccardo Torchio was with the Department of Industrial Engineering, Università degli Studi di Padova. Maximilian Nolte and Sebastian Schöps were with the Computational Electromagnetics group, Technische Universität at Darmstadt. Albert E. Ruehli was with the EMC Laboratory of Missouri University of Science \& Technology.}% <-this % stops a space
\thanks{Manuscript received April 19, 2021; revised August 16, 2021.}}%

\maketitle

\begin{abstract}
This contribution investigates the connection between Isogeometric Analysis (IgA) and the Partial Element Equivalent Circuit (PEEC) method for electrostatic problems. We demonstrate that using the spline-based geometry concepts from IgA allows for extracting circuit elements without a meshing step. Moreover, the proposed IgA-PEEC method converges for complex geometries up to three times faster than the conventional PEEC approach and, in turn, it requires a significantly lower number of degrees of freedom to solve a problem with comparable accuracy. The resulting method is closely related to the isogeometric boundary element method. However, it uses lowest-order basis functions to allow for straightforward physical and circuit interpretations. The findings are validated by an analytical example with complex geometry, i.e., significant curvature, and by a realistic model of a surge arrester.
\end{abstract}

\bigskip


\section{Introduction}
Numerical simulation tools based on discretizations of Maxwell's equations are established in academia and industry. They support engineers during the design and analysis of products like electric surge arresters or high voltage devices \cite{journal5,journal4}. They are particularly popular when designing complex geometries that cannot be analyzed in closed-form or by simple electric circuit models. 

Most such numerical methods are based on surface or volume meshes that approximate the geometry with low order elements, e.g., triangles. This comes with two drawbacks: First, there is a geometry error that is only acceptable if it does not impede the overall convergence of the method. Secondly, the meshing step may be time consuming and error prone. Its contribution to the overall workflow can be significant, Sandia labs have analyzed that about 75\% of the simulation time is related to modeling, parameterization, mesh generation and pre- and post-processing \cite{Boggs_2005aa}. In contrast, Isogeometric Analysis (IgA), \cite{Cottrell_2009aa}, makes use of the exact representation of a device's geometry in the language of Computer Aided Design (CAD) tools. Most common CAD tools use boundary representations (B-rep) of objects based on patches of Non Uniform Rational B-Splines (NURBS). NURBS are popular since they can describe conic sections exactly, enable local smoothness control, and allow defining curves
and surfaces intuitively~\cite{Cohen_2001aa}. 

\begin{figure}[t!]
	\centering
	\begin{tikzpicture}[scale=2.95]
		\node at (0,0) {\includegraphics[width=5.3cm]{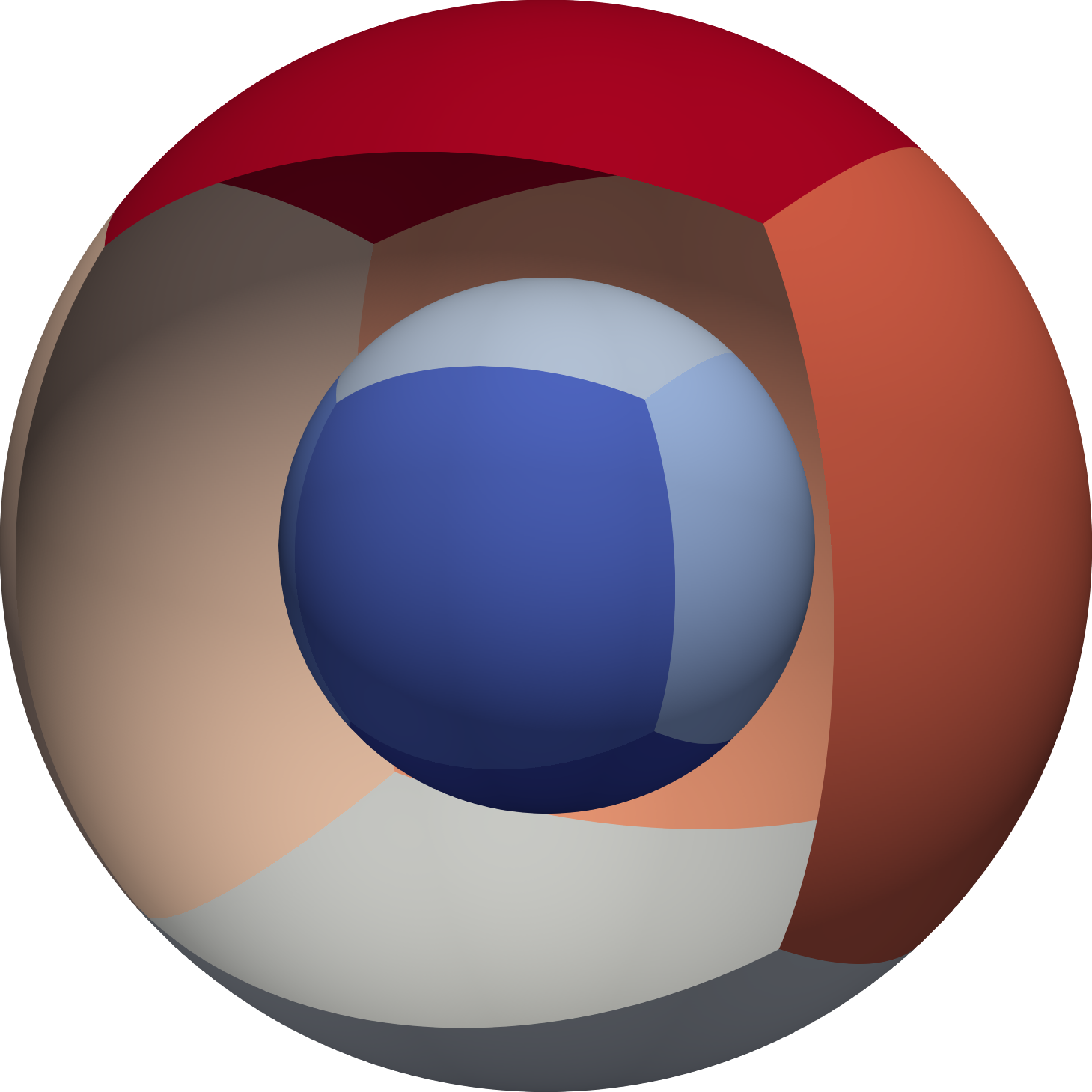}};
		% capacitors
		% bottom
		\def\a{0.7};
		\def\b{0.65};
		\draw [black] (0,-0.9) -- (0,-\a);
		\draw [mark=ball,fill] (0,-0.9) circle [radius=0.17mm];
		\draw [black] (-0.125,-\a) -- (0.125,-\a);
		\draw [black] (-0.125,-\b) -- (0.125,-\b);
		\draw [black] (0,-\b) -- (0,-0.44);
		\draw [mark=ball,fill] (0,-0.44) circle [radius=0.17mm];

		% left infty
		\draw [black] (-0.9,0) -- (-1.1,0);
		\draw [black] (-1.1,-0.125) -- (-1.1,0.125);
		\draw [black] (-1.15,-0.125) -- (-1.15,0.125);
		\draw [black] (-1.15,0) -- (-1.35,0);
		\draw [mark=ball,fill] (-1.35,0) circle [radius=0.17mm];
		\node [anchor=south] at (-1.35,0) {\tiny$\infty$};

		% left
		\draw [black] (-0.9,0) -- (-\a,0);
		\draw [mark=ball,fill] (-0.9,0) circle [radius=0.17mm];
		\draw [black] (-\a,-0.125) -- (-\a,0.125);
		\draw [black] (-\b,-0.125) -- (-\b,0.125);
		\draw [black] (-\b,0) -- (-0.44,0);
		\draw [mark=ball,fill] (-0.44,0) circle [radius=0.17mm];

		% top	
		\draw [black] (0,0.9) -- (0,\a);
		\draw [mark=ball,fill] (0,0.9) circle [radius=0.17mm];
		\draw [black] (-0.125,\a) -- (0.125,\a);
		\draw [black] (-0.125,\b) -- (0.125,\b);
		\draw [black] (0,\b) -- (0,0.44);
		\draw [mark=ball,fill] (0,0.44) circle [radius=0.17mm];

		% right
		\draw [black] (0.9,0) -- (\a,0);
		\draw [mark=ball,fill] (0.9,0) circle [radius=0.17mm];
		\draw [black] (\a,-0.125) -- (\a,0.125);
		\draw [black] (\b,-0.125) -- (\b,0.125);
		\draw [black] (\b,0) -- (0.44,0);
		\draw [mark=ball,fill] (0.44,0) circle [radius=0.17mm];

		% rotated top right
		\def\shift0{0.22cm}
		\begin{scope}[xshift=\shift0,yshift=\shift0,rotate around={-45:(0,0)}]
		\draw [black] (0.025,-0.125) -- (0.025,0.125);
		\draw [black] (-0.025,-0.125) -- (-0.025,0.125);
		\draw [black] (0.025,0) -- (0.3,0);
		\draw [black] (-0.025,0) -- (-0.3,0);
		\end{scope}

		% rotated bottom right
		\begin{scope}[xshift=\shift0,yshift=-\shift0,rotate around={-135:(0,0)}]
		\draw [black] (0.025,-0.125) -- (0.025,0.125);
		\draw [black] (-0.025,-0.125) -- (-0.025,0.125);
		\draw [black] (0.025,0) -- (0.3,0);
		\draw [black] (-0.025,0) -- (-0.3,0);
		\end{scope}

		%rotated bottom left
		\begin{scope}[xshift=-\shift0,yshift=-\shift0,rotate around={-45:(0,0)}]
		\draw [black] (0.025,-0.125) -- (0.025,0.125);
		\draw [black] (-0.025,-0.125) -- (-0.025,0.125);
		\draw [black] (0.025,0) -- (0.3,0);
		\draw [black] (-0.025,0) -- (-0.3,0);
		\end{scope}

		%rotated top left
		\begin{scope}[xshift=-\shift0,yshift=\shift0,rotate around={-135:(0,0)}]
		\draw [black] (0.025,-0.125) -- (0.025,0.125);
		\draw [black] (-0.025,-0.125) -- (-0.025,0.125);
		\draw [black] (0.025,0) -- (0.3,0);
		\draw [black] (-0.025,0) -- (-0.3,0);
		\end{scope}

		% rotated outer top right
		\def\c{0.25}
		\def\shift{.6cm}
		\begin{scope}[xshift=\shift,yshift=\shift,rotate around={-45:(0,0)}]
		\draw [black] (0.025,-0.125) -- (0.025,0.125);
		\draw [black] (-0.025,-0.125) -- (-0.025,0.125);
		\draw [black] (0.025,0) -- (\c,0);
		\draw [black] (-0.025,0) -- (-\c,0);
		\draw [mark=ball,fill] (\c,0) circle [radius=0.17mm];
		\draw [mark=ball,fill] (-\c,0) circle [radius=0.17mm];
		\end{scope}

		% rotated outer bottom right
		\begin{scope}[xshift=\shift,yshift=-\shift,rotate around={-135:(0,0)}]
		\draw [black] (0.025,-0.125) -- (0.025,0.125);
		\draw [black] (-0.025,-0.125) -- (-0.025,0.125);
		\draw [black] (0.025,0) -- (\c,0);
		\draw [black] (-0.025,0) -- (-\c,0);
		\draw [mark=ball,fill] (\c,0) circle [radius=0.17mm];
		\draw [mark=ball,fill] (-\c,0) circle [radius=0.17mm];
		\end{scope}

		% rotated outer bottom left
		\begin{scope}[xshift=-\shift,yshift=-\shift,rotate around={-45:(0,0)}]
		\draw [black] (0.025,-0.125) -- (0.025,0.125);
		\draw [black] (-0.025,-0.125) -- (-0.025,0.125);
		\draw [black] (0.025,0) -- (\c,0);
		\draw [black] (-0.025,0) -- (-\c,0);
		\draw [mark=ball,fill] (\c,0) circle [radius=0.17mm];
		\draw [mark=ball,fill] (-\c,0) circle [radius=0.17mm];
		\end{scope}

		% rotated outer top right
		\begin{scope}[xshift=-\shift,yshift=\shift,rotate around={-135:(0,0)}]
		\draw [black] (0.025,-0.125) -- (0.025,0.125);
		\draw [black] (-0.025,-0.125) -- (-0.025,0.125);
		\draw [black] (0.025,0) -- (\c,0);
		\draw [black] (-0.025,0) -- (-\c,0);
		\draw [mark=ball,fill] (\c,0) circle [radius=0.17mm];
		\draw [mark=ball,fill] (-\c,0) circle [radius=0.17mm];
		\end{scope}
	\end{tikzpicture}
	\caption{Interpretation of the coefficients $C_{k,l}$ as capacities, the remaining capacities are being omitted for the purpose of visual clarity.}
	\label{fig:capa}
\end{figure}
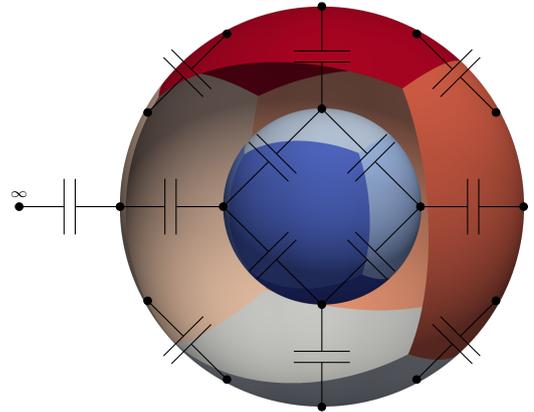

We propose to formulate the electrostatic problem in terms of integral equations based on the same NURBS geometry descriptions used in IgA. Consequently, there is no geometrical approximation error and no need of a separate meshing step. We follow the well-known Partial Element Equivalent Circuit (PEEC) scheme \cite{Ruehli_1974aa}. It allows for an automatic representation of the problem by means of an equivalent electric circuit, see Fig.~\ref{fig:capa}. Eventually, the circuit can be solved in a SPICE-like solver, possibly combined with other circuits that represent other devices and/or parts of the system. For example, nonlinear models that are popular to describe varistor-based stress control of high-voltage surge arresters \cite{Denz_2014aa}, can be easily included. Being in the isogeometric setting, we use the term patch instead of the term cell commonly used in PEEC methods. The usage of the two terms is equivalent.

The original PEEC method \cite{Ruehli_1974aa,Ruehli_2015aa} is based on subdividing the computational domain in simple structures, e.g., regular hexahedra or rectangular patches, that allow exact and fast quadrature of the involved integrals. During the years, the PEEC method has been extended to the case of more general mesh elements, such as non-orthogonal hexahedral elements \cite{1032521}, tetrahedral elements \cite{8764572}, etc. However, in the literature only linear (non-curved) elements and lowest order basis functions have been used in the context of PEEC. Therefore, geometrical approximation errors are unavoidable when dealing with complex freeform geometries, e.g., devices with curvatures \cite{journal1}. 

Our implementation is based on an existing isogeometric boundary element library \cite{Dolz_2018aa,Dolz_2020ac}. It is shown that using this framework with lowest-order basis functions for the expansion of the solution, i.e., one degree of freedom (electric charge) per patch and high-order rational basis functions for the geometry, a PEEC method is obtained that fulfills all promises mentioned above. In particular, it is demonstrated using the example of a spherical capacitor, that the convergence rate of this new method is optimal while the rate deteriorates when using conventional PEEC implementations. This is explained by the fact that the so-called Aubin-Nitsche trick, e.g., \cite{Sauter_2011aa}, is not applicable due to the geometric modelling error.

Consequently, the proposed combination of technologies contributes to bridging the gap between CAD and circuit simulation. The rest of the paper is organized as follows. In Section \ref{sec:splines}, the fundamentals of CAD and NURBS is revisited. Moreover, the specific choice of the shape functions used to apply the PEEC discretization scheme is discussed. In Section \ref{sec:PEEC} the PEEC formulation for the electrostatic case is briefly described, as well as the use of patch-wise
defined B-splines to discretize the unknonws. Section \ref{sec:Stamping} revise the PEEC stamping techniques for the specific case of electrostatic problems, therefore it is shown how a \textit{net-list} can be automatically constructed from the matrix coefficients generated by the IgA-PEEC method. 
Finally, in Section \ref{sec:num_res} the findings are
validated by considering an analytical example with complex geometry,
i.e., significant curvature, and  a realistic model of a surge arrester.

\section{Splines and Geometry} \label{sec:splines}
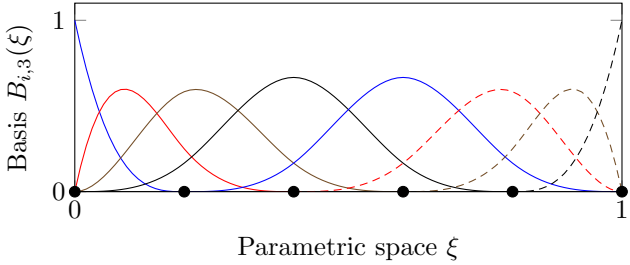
\begin{figure}[t!]
	\centering

	\begin{tikzpicture}

	\begin{axis}[%
	no markers,
	width=0.4\textwidth,
	height=2.5cm,
	scale only axis,
	xlabel near ticks,
	ylabel near ticks,
	xmin=0,
	xmax=1,
	ymin=0,
	ymax=1.1,
	xlabel= {Parametric space $\xi$},
	ylabel= {Basis $B_{i,3}(\xi)$},
	ytick distance=1,
	xtick distance=1,
	axis background/.style={fill=white}
	]
	\addplot table[row sep=crcr]{%
	0	1\\
	0.0105263157894737	0.850269718617874\\
	0.0210526315789474	0.716285172765709\\
	0.0315789473684211	0.597171599358507\\
	0.0421052631578947	0.49205423531127\\
	0.0526315789473684	0.400058317539\\
	0.0631578947368421	0.320309082956699\\
	0.0736842105263158	0.25193176847937\\
	0.0842105263157895	0.194051611022015\\
	0.0947368421052632	0.145793847499636\\
	0.105263157894737	0.106283714827234\\
	0.115789473684211	0.0746464499198134\\
	0.126315789473684	0.050007289692375\\
	0.136842105263158	0.0314914710599213\\
	0.147368421052632	0.0182242309374544\\
	0.157894736842105	0.00933080623997667\\
	0.168421052631579	0.00393643388249015\\
	0.178947368421053	0.00116635077999709\\
	0.189473684210526	0.000145793847499636\\
	0.2	0\\
	};
	\addplot table[row sep=crcr]{%
	0	0\\
	0.0105263157894737	0.145684502114011\\
	0.0210526315789474	0.26796909170433\\
	0.0315789473684211	0.368384604169704\\
	0.0421052631578947	0.448461874908879\\
	0.0526315789473684	0.509731739320601\\
	0.0631578947368421	0.553725032803616\\
	0.0736842105263158	0.58197259075667\\
	0.0842105263157895	0.59600524857851\\
	0.0947368421052632	0.597353841667882\\
	0.105263157894737	0.587549205423531\\
	0.115789473684211	0.568122175244205\\
	0.126315789473684	0.540603586528648\\
	0.136842105263158	0.506524274675609\\
	0.147368421052632	0.467415075083831\\
	0.157894736842105	0.424806823152063\\
	0.168421052631579	0.380230354279049\\
	0.178947368421053	0.335216503863537\\
	0.189473684210526	0.291296107304272\\
	0.2	0.25\\
	0.210526315789474	0.212567429654469\\
	0.221052631578947	0.179071293191427\\
	0.231578947368421	0.149292899839627\\
	0.242105263157895	0.123013558827818\\
	0.252631578947368	0.10001457938475\\
	0.263157894736842	0.0800772707391748\\
	0.273684210526316	0.0629829421198425\\
	0.284210526315789	0.0485129027555037\\
	0.294736842105263	0.0364484618749089\\
	0.305263157894737	0.0265709287068086\\
	0.315789473684211	0.0186616124799533\\
	0.326315789473684	0.0125018224230938\\
	0.336842105263158	0.00787286776498031\\
	0.347368421052632	0.00455605773436361\\
	0.357894736842105	0.00233270155999417\\
	0.368421052631579	0.000984108470622539\\
	0.378947368421053	0.000291587694999271\\
	0.389473684210526	3.64484618749086e-05\\
	0.4	0\\
	};
	\addplot table[row sep=crcr]{%
	0	0\\
	0.0105263157894737	0.00402148029353161\\
	0.0210526315789474	0.0155513437332945\\
	0.0315789473684211	0.0337877241580405\\
	0.0421052631578947	0.0579287554065218\\
	0.0526315789473684	0.0871725713174904\\
	0.0631578947368421	0.120717305729698\\
	0.0736842105263158	0.157761092481897\\
	0.0842105263157895	0.19750206541284\\
	0.0947368421052632	0.239138358361277\\
	0.105263157894737	0.281868105165962\\
	0.115789473684211	0.324889439665646\\
	0.126315789473684	0.367400495699082\\
	0.136842105263158	0.40859940710502\\
	0.147368421052632	0.447684307722214\\
	0.157894736842105	0.483853331389415\\
	0.168421052631579	0.516304611945376\\
	0.178947368421053	0.544236283228848\\
	0.189473684210526	0.566846479078583\\
	0.2	0.583333333333333\\
	0.210526315789474	0.593113670603101\\
	0.221052631578947	0.596479078582884\\
	0.231578947368421	0.593939835738932\\
	0.242105263157895	0.586006220537493\\
	0.252631578947368	0.573188511444817\\
	0.263157894736842	0.555996986927152\\
	0.273684210526316	0.534941925450746\\
	0.284210526315789	0.510533605481849\\
	0.294736842105263	0.483282305486708\\
	0.305263157894737	0.453698303931574\\
	0.315789473684211	0.422291879282694\\
	0.326315789473684	0.389573310006318\\
	0.336842105263158	0.356052874568693\\
	0.347368421052632	0.322240851436069\\
	0.357894736842105	0.288647519074695\\
	0.368421052631579	0.255783155950819\\
	0.378947368421053	0.22415804053069\\
	0.389473684210526	0.194282451280556\\
	0.4	0.166666666666667\\
	0.410526315789474	0.141711619769646\\
	0.421052631578947	0.119380862127618\\
	0.431578947368421	0.0995285998930845\\
	0.442105263157895	0.0820090392185449\\
	0.452631578947368	0.0666763862564999\\
	0.463157894736842	0.0533848471594499\\
	0.473684210526316	0.041988628079895\\
	0.484210526315789	0.0323419351703358\\
	0.494736842105263	0.0242989745832726\\
	0.505263157894737	0.0177139524712057\\
	0.515789473684211	0.0124410749866355\\
	0.526315789473684	0.0083345482820625\\
	0.536842105263158	0.00524857850998687\\
	0.547368421052632	0.00303737182290906\\
	0.557894736842105	0.00155513437332945\\
	0.568421052631579	0.000656072313748359\\
	0.578947368421053	0.000194391796666179\\
	0.589473684210526	2.42989745832728e-05\\
	0.6	0\\
	};
	\addplot table[row sep=crcr]{%
	0	0\\
	0.0105263157894737	2.42989745832726e-05\\
	0.0210526315789474	0.000194391796666181\\
	0.0315789473684211	0.00065607231374836\\
	0.0421052631578947	0.00155513437332945\\
	0.0526315789473684	0.00303737182290907\\
	0.0631578947368421	0.00524857850998688\\
	0.0736842105263158	0.0083345482820625\\
	0.0842105263157895	0.0124410749866356\\
	0.0947368421052632	0.0177139524712057\\
	0.105263157894737	0.0242989745832726\\
	0.115789473684211	0.0323419351703358\\
	0.126315789473684	0.041988628079895\\
	0.136842105263158	0.0533848471594499\\
	0.147368421052632	0.0666763862565\\
	0.157894736842105	0.082009039218545\\
	0.168421052631579	0.0995285998930846\\
	0.178947368421053	0.119380862127618\\
	0.189473684210526	0.141711619769646\\
	0.2	0.166666666666667\\
	0.210526315789474	0.194294600767848\\
	0.221052631578947	0.224255236429023\\
	0.231578947368421	0.256111192107693\\
	0.242105263157895	0.28942508626136\\
	0.252631578947368	0.323759537347524\\
	0.263157894736842	0.358677163823687\\
	0.273684210526316	0.393740584147349\\
	0.284210526315789	0.428512416776012\\
	0.294736842105263	0.462555280167177\\
	0.305263157894737	0.495431792778345\\
	0.315789473684211	0.526704573067017\\
	0.326315789473684	0.555936239490693\\
	0.336842105263158	0.582689410506877\\
	0.347368421052632	0.606526704573067\\
	0.357894736842105	0.627010740146766\\
	0.368421052631579	0.643704135685474\\
	0.378947368421053	0.656169509646693\\
	0.389473684210526	0.663969480487923\\
	0.4	0.666666666666667\\
	0.410526315789474	0.663969480487923\\
	0.421052631578947	0.656169509646693\\
	0.431578947368421	0.643704135685474\\
	0.442105263157895	0.627010740146766\\
	0.452631578947368	0.606526704573067\\
	0.463157894736842	0.582689410506877\\
	0.473684210526316	0.555936239490693\\
	0.484210526315789	0.526704573067016\\
	0.494736842105263	0.495431792778345\\
	0.505263157894737	0.462555280167177\\
	0.515789473684211	0.428512416776012\\
	0.526315789473684	0.393740584147349\\
	0.536842105263158	0.358677163823687\\
	0.547368421052632	0.323759537347524\\
	0.557894736842105	0.28942508626136\\
	0.568421052631579	0.256111192107693\\
	0.578947368421053	0.224255236429023\\
	0.589473684210526	0.194294600767848\\
	0.6	0.166666666666667\\
	0.610526315789474	0.141711619769646\\
	0.621052631578947	0.119380862127618\\
	0.631578947368421	0.0995285998930846\\
	0.642105263157895	0.082009039218545\\
	0.652631578947368	0.0666763862565\\
	0.663157894736842	0.0533848471594499\\
	0.673684210526316	0.0419886280798951\\
	0.684210526315789	0.0323419351703358\\
	0.694736842105263	0.0242989745832726\\
	0.705263157894737	0.0177139524712057\\
	0.715789473684211	0.0124410749866356\\
	0.726315789473684	0.0083345482820625\\
	0.736842105263158	0.00524857850998687\\
	0.747368421052632	0.00303737182290908\\
	0.757894736842105	0.00155513437332945\\
	0.768421052631579	0.000656072313748359\\
	0.778947368421053	0.000194391796666179\\
	0.789473684210526	2.42989745832728e-05\\
	0.8	0\\
	};
	\addplot table[row sep=crcr]{%
	0.2	0\\
	0.210526315789474	2.42989745832726e-05\\
	0.221052631578947	0.000194391796666181\\
	0.231578947368421	0.000656072313748359\\
	0.242105263157895	0.00155513437332944\\
	0.252631578947368	0.00303737182290907\\
	0.263157894736842	0.00524857850998689\\
	0.273684210526316	0.0083345482820625\\
	0.284210526315789	0.0124410749866356\\
	0.294736842105263	0.0177139524712057\\
	0.305263157894737	0.0242989745832726\\
	0.315789473684211	0.0323419351703358\\
	0.326315789473684	0.041988628079895\\
	0.336842105263158	0.0533848471594499\\
	0.347368421052632	0.0666763862565\\
	0.357894736842105	0.082009039218545\\
	0.368421052631579	0.0995285998930846\\
	0.378947368421053	0.119380862127618\\
	0.389473684210526	0.141711619769646\\
	0.4	0.166666666666667\\
	0.410526315789474	0.194294600767847\\
	0.421052631578947	0.224255236429023\\
	0.431578947368421	0.256111192107693\\
	0.442105263157895	0.28942508626136\\
	0.452631578947368	0.323759537347524\\
	0.463157894736842	0.358677163823687\\
	0.473684210526316	0.393740584147349\\
	0.484210526315789	0.428512416776012\\
	0.494736842105263	0.462555280167177\\
	0.505263157894737	0.495431792778345\\
	0.515789473684211	0.526704573067017\\
	0.526315789473684	0.555936239490693\\
	0.536842105263158	0.582689410506877\\
	0.547368421052632	0.606526704573067\\
	0.557894736842105	0.627010740146766\\
	0.568421052631579	0.643704135685474\\
	0.578947368421053	0.656169509646693\\
	0.589473684210526	0.663969480487923\\
	0.6	0.666666666666667\\
	0.610526315789474	0.663969480487923\\
	0.621052631578947	0.656169509646693\\
	0.631578947368421	0.643704135685474\\
	0.642105263157895	0.627010740146766\\
	0.652631578947368	0.606526704573067\\
	0.663157894736842	0.582689410506877\\
	0.673684210526316	0.555936239490693\\
	0.684210526315789	0.526704573067016\\
	0.694736842105263	0.495431792778345\\
	0.705263157894737	0.462555280167177\\
	0.715789473684211	0.428512416776012\\
	0.726315789473684	0.393740584147349\\
	0.736842105263158	0.358677163823686\\
	0.747368421052632	0.323759537347524\\
	0.757894736842105	0.28942508626136\\
	0.768421052631579	0.256111192107693\\
	0.778947368421053	0.224255236429022\\
	0.789473684210526	0.194294600767848\\
	0.8	0.166666666666667\\
	0.810526315789474	0.141711619769646\\
	0.821052631578947	0.119380862127618\\
	0.831578947368421	0.0995285998930845\\
	0.842105263157895	0.0820090392185449\\
	0.852631578947368	0.0666763862565\\
	0.863157894736842	0.0533848471594499\\
	0.873684210526316	0.041988628079895\\
	0.88421052631579	0.0323419351703357\\
	0.894736842105263	0.0242989745832726\\
	0.905263157894737	0.0177139524712057\\
	0.91578947368421	0.0124410749866356\\
	0.926315789473684	0.0083345482820625\\
	0.936842105263158	0.00524857850998687\\
	0.947368421052632	0.00303737182290906\\
	0.957894736842105	0.00155513437332945\\
	0.968421052631579	0.000656072313748359\\
	0.978947368421053	0.000194391796666179\\
	0.989473684210526	2.42989745832728e-05\\
	1	0\\
	};
	\addplot table[row sep=crcr]{%
	0.4	0\\
	0.410526315789474	2.42989745832724e-05\\
	0.421052631578947	0.000194391796666181\\
	0.431578947368421	0.000656072313748359\\
	0.442105263157895	0.00155513437332945\\
	0.452631578947368	0.00303737182290907\\
	0.463157894736842	0.00524857850998687\\
	0.473684210526316	0.0083345482820625\\
	0.484210526315789	0.0124410749866356\\
	0.494736842105263	0.0177139524712057\\
	0.505263157894737	0.0242989745832726\\
	0.515789473684211	0.0323419351703358\\
	0.526315789473684	0.041988628079895\\
	0.536842105263158	0.0533848471594499\\
	0.547368421052632	0.0666763862565\\
	0.557894736842105	0.0820090392185449\\
	0.568421052631579	0.0995285998930845\\
	0.578947368421053	0.119380862127618\\
	0.589473684210526	0.141711619769646\\
	0.6	0.166666666666667\\
	0.610526315789474	0.194282451280556\\
	0.621052631578947	0.224158040530689\\
	0.631578947368421	0.255783155950819\\
	0.642105263157895	0.288647519074695\\
	0.652631578947368	0.32224085143607\\
	0.663157894736842	0.356052874568693\\
	0.673684210526316	0.389573310006318\\
	0.684210526315789	0.422291879282694\\
	0.694736842105263	0.453698303931574\\
	0.705263157894737	0.483282305486708\\
	0.715789473684211	0.510533605481849\\
	0.726315789473684	0.534941925450746\\
	0.736842105263158	0.555996986927152\\
	0.747368421052632	0.573188511444817\\
	0.757894736842105	0.586006220537493\\
	0.768421052631579	0.593939835738932\\
	0.778947368421053	0.596479078582884\\
	0.789473684210526	0.5931136706031\\
	0.8	0.583333333333333\\
	0.810526315789474	0.566846479078583\\
	0.821052631578947	0.544236283228848\\
	0.831578947368421	0.516304611945376\\
	0.842105263157895	0.483853331389415\\
	0.852631578947368	0.447684307722214\\
	0.863157894736842	0.40859940710502\\
	0.873684210526316	0.367400495699081\\
	0.88421052631579	0.324889439665646\\
	0.894736842105263	0.281868105165962\\
	0.905263157894737	0.239138358361277\\
	0.91578947368421	0.19750206541284\\
	0.926315789473684	0.157761092481897\\
	0.936842105263158	0.120717305729698\\
	0.947368421052632	0.0871725713174902\\
	0.957894736842105	0.0579287554065219\\
	0.968421052631579	0.0337877241580405\\
	0.978947368421053	0.0155513437332944\\
	0.989473684210526	0.00402148029353164\\
	1	0\\
	};
	\addplot table[row sep=crcr]{%
	0.6	0\\
	0.610526315789474	3.64484618749092e-05\\
	0.621052631578947	0.000291587694999269\\
	0.631578947368421	0.000984108470622538\\
	0.642105263157895	0.00233270155999417\\
	0.652631578947368	0.00455605773436362\\
	0.663157894736842	0.00787286776498031\\
	0.673684210526316	0.0125018224230937\\
	0.684210526315789	0.0186616124799534\\
	0.694736842105263	0.0265709287068086\\
	0.705263157894737	0.0364484618749088\\
	0.715789473684211	0.0485129027555037\\
	0.726315789473684	0.0629829421198426\\
	0.736842105263158	0.0800772707391749\\
	0.747368421052632	0.10001457938475\\
	0.757894736842105	0.123013558827818\\
	0.768421052631579	0.149292899839627\\
	0.778947368421053	0.179071293191428\\
	0.789473684210526	0.212567429654469\\
	0.8	0.25\\
	0.810526315789474	0.291296107304272\\
	0.821052631578947	0.335216503863537\\
	0.831578947368421	0.380230354279049\\
	0.842105263157895	0.424806823152063\\
	0.852631578947368	0.467415075083831\\
	0.863157894736842	0.506524274675609\\
	0.873684210526316	0.540603586528649\\
	0.88421052631579	0.568122175244205\\
	0.894736842105263	0.587549205423531\\
	0.905263157894737	0.597353841667882\\
	0.91578947368421	0.59600524857851\\
	0.926315789473684	0.58197259075667\\
	0.936842105263158	0.553725032803616\\
	0.947368421052632	0.5097317393206\\
	0.957894736842105	0.448461874908879\\
	0.968421052631579	0.368384604169704\\
	0.978947368421053	0.26796909170433\\
	0.989473684210526	0.145684502114011\\
	1	0\\
	};
	\addplot table[row sep=crcr]{%
	0.8	0\\
	0.810526315789474	0.000145793847499637\\
	0.821052631578947	0.00116635077999708\\
	0.831578947368421	0.00393643388249016\\
	0.842105263157895	0.00933080623997669\\
	0.852631578947368	0.0182242309374544\\
	0.863157894736842	0.0314914710599213\\
	0.873684210526316	0.050007289692375\\
	0.88421052631579	0.0746464499198135\\
	0.894736842105263	0.106283714827234\\
	0.905263157894737	0.145793847499636\\
	0.91578947368421	0.194051611022015\\
	0.926315789473684	0.25193176847937\\
	0.936842105263158	0.320309082956699\\
	0.947368421052632	0.400058317539\\
	0.957894736842105	0.49205423531127\\
	0.968421052631579	0.597171599358507\\
	0.978947368421053	0.71628517276571\\
	0.989473684210526	0.850269718617874\\
	1	1\\
	};

	\addplot[color=black] coordinates {
	(0,0)
	(1,0)
	};
	\addplot[only marks,mark=*] coordinates {
	(0,0)
	(0.2,0)
	(0.4,0)
	(0.6,0)
	(0.8,0)
	(1,0)
	};
	\end{axis}
	\end{tikzpicture}%
	\caption{B-Splines basis functions of degree $3$.}
	\label{fig:bsline}
\end{figure}
We assume that the CAD geometry is represented by the union of $N_{\Gamma}$ surface patches given in terms of NURBS. Each patch is constructed from two (weighted) b-spline curves. The basis $\{ B_{i,p} \}_{i=1}^{N_1}$ of a one-dimensional B-spline space
$\mathbb{S}^{p}_{\alpha}$ of degree $p$ and regularity $\alpha$ is
determined by the knot vector $\boldsymbol{\Xi} = (\xi, \xi', \dots,
\xi_{n}) \in [0,1]^n$, $\xi \leq \xi' \leq \dots \leq \xi_n$. The basis is then given by the Cox-de Boor algorithm%~\cite{de-Boor_2001aa}
\begin{align}
	B_{i,0}(\xi) &= \begin{cases}
		1 \quad \mathrm{if} \quad \xi_i \leq \xi < \xi_{i+1}\\
		0 \quad \mathrm{otherwise}
	\end{cases}
\end{align}
and for $p>0$
\begin{align*}
	B_{i,p}(\xi) &= \frac{\xi - \xi_i}{\xi_{i+p} - \xi_i} B_{i,p-1}(\xi) +
	\frac{\xi_{i+p+1} - \xi}{\xi_{i+p+1} - \xi_{i+1}} B_{i+1,p-1}(\xi).
\end{align*}
Fig.~\ref{fig:bsline} shows a visualization of the basis functions in of degree $p=3$.
A corresponding NURBS curve is obtained by 
\begin{figure}[t!]
	\centering
	\includegraphics[trim={5mm 5mm 5mm 5mm},clip,width=0.5\linewidth]{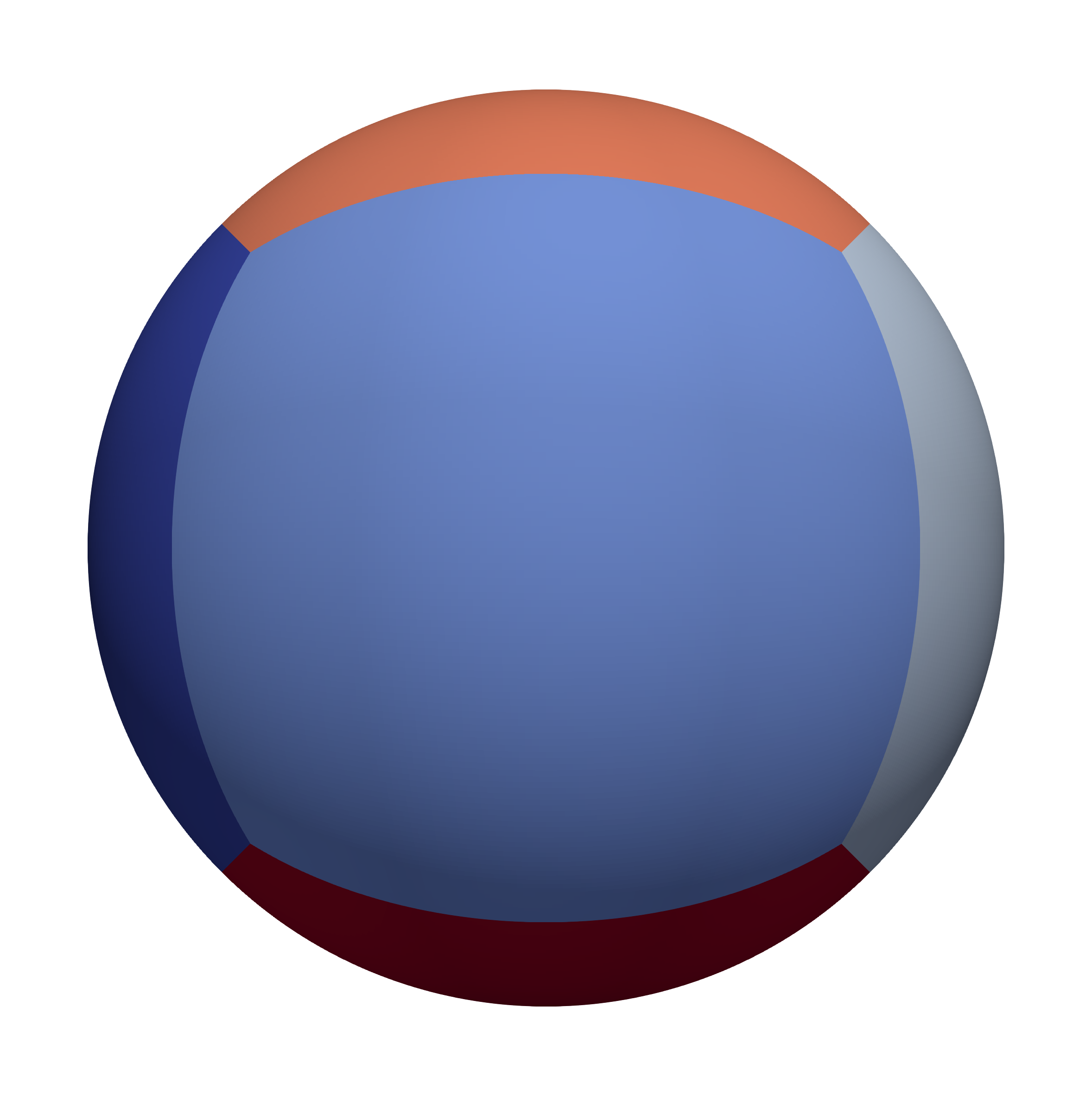}
	\hspace{-1em}
	\includegraphics[trim=45mm 28mm 45mm 28mm, clip,width=0.5\columnwidth]{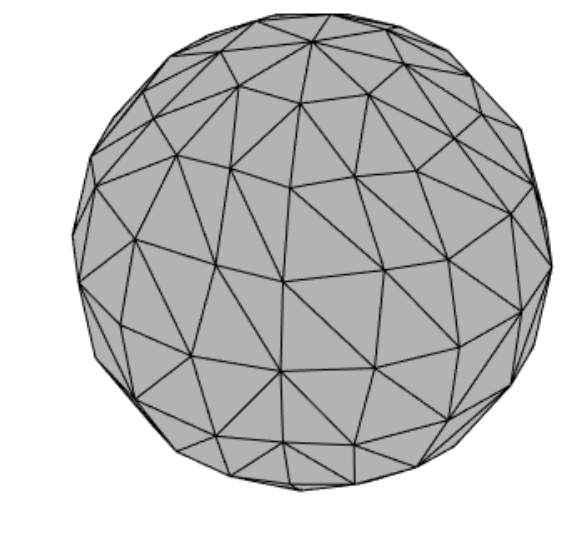}
	\caption{Exact description of a sphere by $N_{\Gamma}=6$ NURBS patches (left) and approximation of a sphere by 208 triangles (right).}
	\label{fig:sphere_exact}
\end{figure}
\begin{align}
\gamma(\xi) = \sum_{i=1}^k\frac
	{{B_{i,p}(\xi)w_i \mathbf{p}_i}}
	{\sum_{q=1}^k {B_{q,p}(\xi)w_q}}
\end{align}
where $k$ is the number of control points $\mathbf{p}_i$ and $w_i$ are the weights. Finally, the $n$-th patch can be described as a NURBS mapping form the reference space $[0,1]\times[0,1]$ to the three-dimensional physical space as
\begin{align}
	\Gamma_n(\xi,\xi') = \sum_{i=1}^k \sum_{j=1}^l \frac { B_{i,n}(\xi) B_{j,m}(\xi') w_{i,j}\mathbf{p}_{i,j}}{\sum_{q=1}^k \sum_{r=1}^l B_{q,n}(\xi) B_{r,m}(\xi') w_{q,r}}.
\end{align}
Consequently, each distinct conductive domain in the problem is exactly described by the union of a connected set of such patches. Since we are focusing on electrostatic problems, the connected patches making up a domain will be equipotential. We also refer to them as electrodes.

In contrast to many other approaches, only a few patches are sufficient to describe even complex objects, e.g., $N_{\Gamma}=6$ NURBS surfaces determine exactly a sphere, see Fig.~\ref{fig:sphere_exact} (left). Now, when computing integrals on such a surface, e.g., the area $\int_{\Gamma}\mathrm{d}\Gamma'$, then one can use patch-wise Gaussian quadrature on the reference domain using push-forwards \cite{Buffa_2010aa}. The convergence is only determined by the order of the quadrature rule. In contrast, each element of a low-order surface mesh, e.g., shown in Fig.~\ref{fig:sphere_exact} (right), can be exactly evaluated but is limited to low-order convergence. Please note that the quadrature in both cases can be easily parallelized.

\section{Electrostatic PEEC Method} \label{sec:PEEC}
The electrostatic PEEC formulation \cite[eq: (3.52)]{Ruehli_2015aa} starts from the well-known electrostatic field equations:
\begin{equation} \label{eq:divD=rhoE}
	\nabla \cdot \mathbf{D}=\rho, 
\end{equation}
\begin{equation} \label{eq:E=-gradPhi} \mathbf{E}=-\nabla\varphi, 
\end{equation}
where $\varphi$ is the scalar electric potential, $\rho$ is the charge density, $\mathbf{E}$ is the electric field, and $\mathbf{D}$ is the electric displacement field.

For the case where conductive media and free space are considered, $\mathbf{D}$ and $\mathbf{E}$ are linked by the following constitutive equation
\begin{equation} \label{eq:D=eE}
	\mathbf{D}=\varepsilon_0 \mathbf{E},
\end{equation}
where $\varepsilon_0$ is the vacuum permittivity.

Substituting  \eqref{eq:D=eE} into  \eqref{eq:divD=rhoE} and using \eqref{eq:E=-gradPhi} the following Poisson equation is obtained 
\begin{equation} \label{eq:poisson}
	\nabla\cdot\mathbf{E}=-\nabla^2\varphi=\frac{\rho}{\varepsilon_0}.
\end{equation}

The Poisson equation \eqref{eq:poisson} can then be reformulated in terms of Green function $g(\mathbf{r}, \mathbf{r}^{\prime})$, i.e.,
the solution of
\begin{equation} \label{eq:dirac}
	\nabla^2 g(\mathbf{r}, \mathbf{r}^{\prime})=-\delta(\mathbf{r}-\mathbf{r}^\prime),
\end{equation}
where $\mathbf{r}$ is the field point, $\mathbf{r}^\prime$ is the integration point,  $\delta$ is the Dirac delta function, and 
\begin{equation} \label{eq:green}
   g(\mathbf{r}, \mathbf{r}^{\prime})=\frac{1}{4\pi||\mathbf{r}-\mathbf{r}^\prime||}. 
\end{equation}

By using \eqref{eq:dirac} and \eqref{eq:green} into \eqref{eq:poisson}, the integral solution of the Poisson equation is obtained 
\begin{equation} \label{eq:phie}
	\varphi(\mathbf{r}) = \frac{1}{4\pi\varepsilon_0} \int_{\Gamma} \frac{\rho(\mathbf{r}')}{||\mathbf{r}-\mathbf{r}'||}\,\mathrm{d}\Gamma',
	\end{equation}
where $\Gamma$ is the boundary of the conductive domains in terms of NURBS.

For the representation of the fields we may use patch-wise defined B-splines from the space
\begin{equation}
	\mathbb{S}^p_\alpha(\Gamma) = \left\{f\colon f|_{\Gamma_k} \in \mathbb{S}^p_\alpha(\Gamma_k), 1\leq k \leq N_\Gamma \right\},
\end{equation}
with $\mathbb{S}^p_\alpha(\Gamma_k)=\{f\colon(f \circ \Gamma_k) \in \mathbb{S}^p_\alpha\times\mathbb{S}^p_\alpha\}$, see \cite{Buffa_2019ac}.
The goal is to find a discrete solution $\rho_h \in \mathbb{S}^p_\alpha(\Gamma)$, which yields the applied potential $\varphi$ when inserted into \eqref{eq:phie} and evaluated on the electrodes.
For this we choose a basis $\{w_i\}_{1\leq i \leq n}$, with $n=\dim\left(\mathbb{S}^p_\alpha(\Gamma)\right)$ which fulfills $\spanS\{w_1,\dots, w_n\} = \mathbb{S}^p_\alpha(\Gamma)$.
In the particular case of $p=0$ and without mesh refinement, i.e., $i=k$, we choose $w_k$ such that $\hat{w}_k = (w_k \circ \Gamma_k) = 1/|\Gamma_k|$ holds, where $|\Gamma_k|$ denotes the area of the corresponding patch $\Gamma_k$.
Utilizing the Galerkin discretization $\rho_h = \sum_{i=1}^n q_i w_i$ yields the discrete variational problem: find $\rho_h \in \mathbb{S}^p_\alpha(\Gamma)$ such that
\begin{equation}\label{eq:galerkin}
	\frac{1}{4\pi\varepsilon_0}\int_\Gamma w_j(\mathbf{r})\int_\Gamma \frac{\rho_h(\mathbf{r}')}{||\mathbf{r}-\mathbf{r}'||}\,\mathrm{d}\Gamma'\,\mathrm{d}\Gamma = \int_\Gamma \varphi(\mathbf{r}) w_j(\mathbf{r})\,\mathrm{d}\Gamma,
\end{equation}
for all test functions $w_j\in \mathbb{S}^p_\alpha(\Gamma)$.

\begin{figure} [t!]
	\centering
	\includegraphics[trim=0mm 0mm 0mm 0mm, clip,width = 0.75\columnwidth]{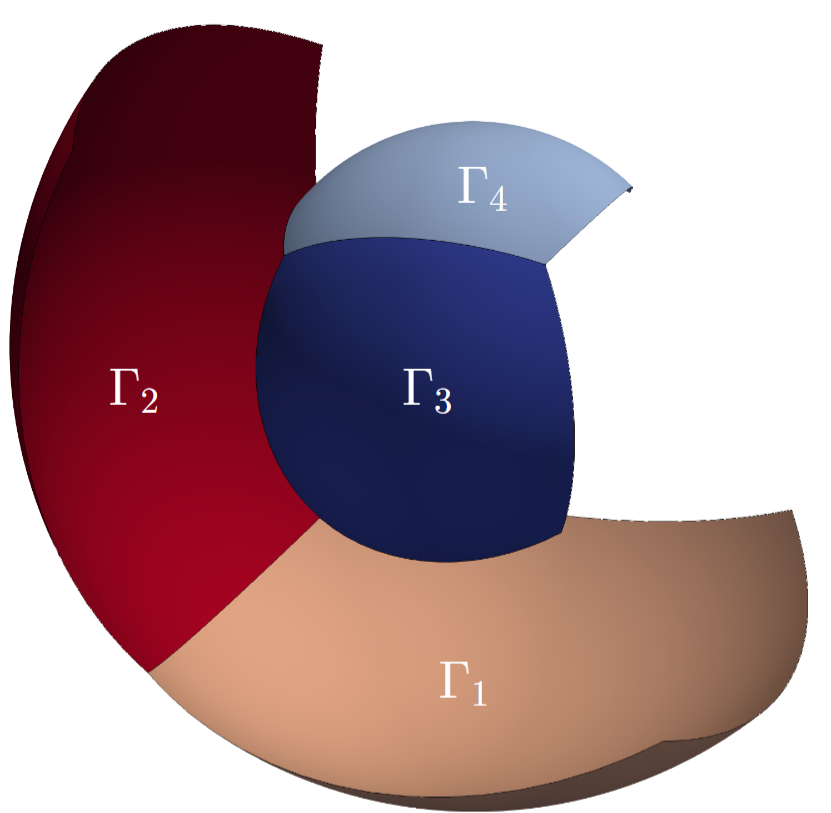}
	\caption{NURBS patches for a simple case with two domains. Each domain is discretized by two patches.}
	\label{Fig.patches}
\end{figure}

\subsection{Quadrature and Solution of the Linear System} 
For the assembly of the Galerkin system matrix \eqref{eq:galerkin}, we need to deal with singular integrals and use the Duffy trick \cite{Duffy_1982aa}.
Additionally, one has to increase the quadrature degree logarithmically with the distance between the evaluated elements \cite{Sauter_2011aa}.
Furthermore, in the utilized implementation Bembel \cite{Dolz_2020ac}, the assembly is performed in parallel.

\begin{figure*}[!t]
	\centering
	\begin{circuitikz}[font=\sffamily, american voltages]
		\draw (0,2.5) to [short,-o] (0.9,2.5);
		\draw (0.9,2.5) to [short,o-] (1.8,2.5);
		\draw (0,2.5) to [short,-] (0,1);
		\draw (1.8,2.5) to [short,-] (1.8,1);
		\draw (0,2.5) to [short,i=$i\omega q_1$] (0,1);
		\draw (1.8,2.5) to [short,i=$i\omega q_2$] (1.8,1);
		\draw (0,1) to [C,-,l=${P}^{-1}_{11}$] (0,0);
		\draw (1.8,0) to [C,-] (1.8,1);		
		\draw (0,0) to (0,0) node[ground]{}; 
		\draw (1.8,0) to (1.8,0) node[ground]{}; 
		\draw (-1,0) to [american controlled current
		source,-,l=$\sum\limits^{4}_{s\neq 1}\frac{{P}_{1s}}{{P}_{11}}i\omega{q}_s$] (-1,1);
		\draw (2.8,0) to [american controlled current
		source,-] (2.8,1);		
		\draw (-1,0) to (-1,0) node[ground]{}; 
		\draw (2.8,0) to (2.8,0) node[ground]{};
		\draw (-1,1.5) to [short,-] (0,1.5);	  
		\draw (1.8,1.5) to [short,-] (2.8,1.5);	 
		\draw (-1,1.5) to [short,-] (-1,1);	  
		\draw (2.8,1.5) to [short,-] (2.8,1);	
		\draw 
		(0.9,2.5) 
		node[label={[font=\footnotesize]above:$n_1$}] {};
		%%%%%%%%%%%%%%%%%%%%
		\draw (5,2.5) to [short,-o] (5.9,2.5);
		\draw (5.9,2.5) to [short,o-] (6.8,2.5);
		\draw (5,2.5) to [short,-] (5,1);
		\draw (6.8,2.5) to [short,-] (6.8,1);
		\draw (5,2.5) to [short,i=$i\omega q_3$] (5,1);
		\draw (6.8,2.5) to [short,i=$i\omega q_4$] (6.8,1);
		\draw (5,1) to [C,-] (5,0);
		\draw (6.8,0) to [C,-] (6.8,1);		
		\draw (5,0) to (5,0) node[ground]{}; 
		\draw (6.8,0) to (6.8,0) node[ground]{}; 
		\draw (4,0) to [american controlled current
		source,-] (4,1);
		\draw (7.8,0) to [american controlled current
		source,-] (7.8,1);		
		\draw (4,0) to (4,0) node[ground]{}; 
		\draw (7.8,0) to (7.8,0) node[ground]{};
		\draw (4,1.5) to [short,-] (5,1.5);	  
		\draw (6.8,1.5) to [short,-] (7.8,1.5);	 
		\draw (4,1.5) to [short,-] (4,1);	  
		\draw (7.8,1.5) to [short,-] (7.8,1);	
		\draw 
		(5.9,2.5) 
		node[label={[font=\footnotesize]above:$n_2$}] {};
	\end{circuitikz}
	\caption{Equivalent circuit  obtained by exploiting the electrostatic PEEC scheme related to the case represented in Fig.~\ref{Fig.patches}.} \label{fig:PEEC_branch}
\end{figure*}
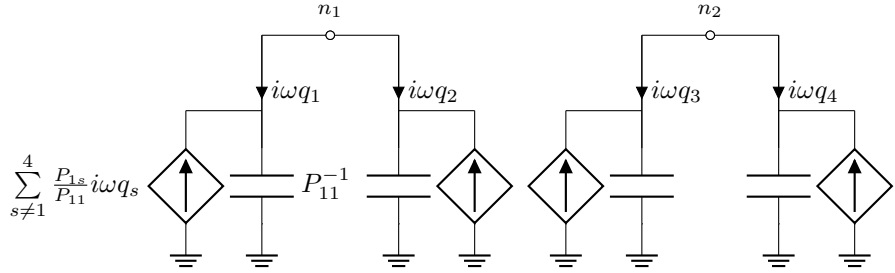

\begin{figure*}[!t]
	\centering
	\begin{circuitikz}[font=\sffamily, american voltages]
		\draw (-1,0) to [american controlled current
		source,-,l=$\sum\limits^{4}_{s\neq 1}\frac{{P}_{1s}}{{P}_{11}}i\omega{q}_s$] (-1,1);
		\draw (-1,1) to [short,-] (-1,1.5);
		\draw (-1,0) to [short,-] (-1,-0.5);
		\draw (-1,-0.5) to (-1,-0.5) node[ground]{};
		\draw (-1,1.5) to [short,-] (0,1.5);
		\draw (0,2) to [short,o-] (0,1.5);
		\draw (0,1.5) to [C,-,l=${P}^{-1}_{11}$] (0,-0.5);
		%\draw (1,-0.5) to [short,-] (1,-1);
		\draw (0,-0.5) to (0,-0.5) node[ground]{};
		\draw 
		(1.75,0.25) 
		node[label={[font=\footnotesize]above:\contour{black}{$\Rightarrow$}}] {};
		\draw (4,0) to [american controlled current
		source,-,l=$\frac{{P}_{12}}{{P}_{11}}i\omega{q}_2$] (4,1);
		\draw (4,1) to [short,-] (4,2);
		\draw (4,0) to [short,-] (4,-1);
		\draw (6.5,0) to [american controlled current
		source,-,l=$\frac{{P}_{13}}{{P}_{11}}i\omega{q}_3$] (6.5,1);
		\draw (6.5,1) to [short,-] (6.5,2);
		\draw (6.5,0) to [short,-] (6.5,-1);
		\draw (6.5,-1) to (6.5,-1) node[ground]{};
		\draw (9,0) to [american controlled current
		source,-,l=$\frac{{P}_{14}}{{P}_{11}}i\omega{q}_4$] (9,1);
		\draw (9,1) to [short,-] (9,2);
		\draw (9,0) to [short,-] (9,-1);
		\draw (4,2) to [short,-] (6.5,2);
		\draw (6.5,2) to [short,-] (9,2);
		\draw (4,-1) to [short,-] (6.5,-1);
		\draw (6.5,-1) to [short,-] (9,-1);
		\draw (11,-1) to [C,-,l=$P_{11}^{-1}$] (11,2);
		\draw (11,-1) to (11,-1) node[ground]{};
		\draw (9,2) to [short,-] (11,2);
		\draw (11,4) to [V,o-,l=0 V] (11,2);
		\draw (11,2) to [short,-,i=$i\omega q_1$] (11,2.5);
	\end{circuitikz}
	\caption{{Representation of the current controlled current sources.}} \label{fig:PEEC_CCCS}
\end{figure*}
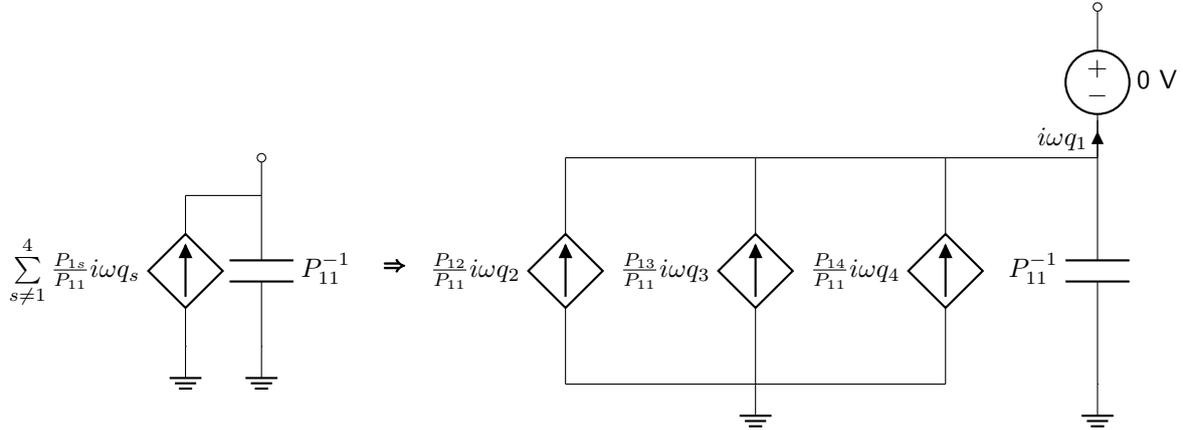

After numerical quadrature, the problem \eqref{eq:galerkin} corresponds to a linear equation system
\begin{equation} \label{eq:Pq=phi}
	\mathbf{P}\mathbf{q}=\pmb{\phi},
\end{equation}
where $\mathbf{q}$ is the coefficient vector describing the charge distribution $\rho_h$,  $\mathbf{P}$ is the potential IgA-PEEC matrix  and $\pmb{\phi}$ contains the potentials. The general $ij$th coefficient of such matrix is given by
\begin{equation}\label{eq:Pcoeff}
	P_{ij}=\frac{1}{4\pi\varepsilon_0}\int_\Gamma w_j(\mathbf{r})\int_\Gamma \frac{w_i(\mathbf{r}')}{||\mathbf{r}-\mathbf{r}'||}\,\mathrm{d}\Gamma'\,\mathrm{d}\Gamma.
\end{equation}
In contrast to (high-order) isogeometric boundary elements methods, we chose element-wise constant basis functions ($p=0$). In the simplest case, i.e., without mesh refinement, the entries of $\mathbf{q}$ can be interpreted as patch-wise charges. Consequently, $\mathbf{C}_M=\mathbf{P}^{-1}$ can be interpreted as the (full) Maxwell capacitance matrix  from which the short circuit capacitance matrix $\mathbf{C}$ can be easily obtained as shown in \cite{Ruehli_2015aa}, where $C_{kl}$, with $k\neq l$, is a capacitance connecting the patches $\Gamma_k$ and $\Gamma_l$, see Fig.~\ref{fig:capa}. Finding  $\mathbf{C}$ requires the inversion of matrix $\mathbf{P}$, that can be computationally expensive. Moreover, as discussed in \cite{Ruehli_2015aa}, some meshing related accuracy problems may occur when matrix  $\mathbf{C}$ want to be obtained from $\mathbf{P}$. For this reason and in the same fashion of the standard PEEC method, with the aim of automatically construct an equivalent electric circuit of \eqref{eq:Pq=phi} that can be loaded and solved into a SPICE-like solver, the current source-based  stamping technique 
described in the next section can be adopted instead. 

\subsection{Stamping Technique} \label{sec:Stamping}
The electrostatic IgA-PEEC problem can be solved with any standard technique for solving linear systems (e.g., LU decomposition). Alternatively, by exploiting the circuit interpretation provided by the PEEC scheme, an equivalent circuit and a \textit{net-list}  can be constructed and the problem can be imported into a SPICE-like solver where it can be solved and  possibly coupled with other circuits representing other devices or parts of the system. In this section, the stamping technique which allows to automatically construct an equivalent \textit{net-list} of the problem is changed for the specific case of electrostatic problems. 

In some applications, it can be convenient to represent the fully coupled electrostatic problem \eqref{eq:Pq=phi} in terms of an equivalent circuit. Thus, a \textit{net-list} which represents the electrostatics problem can be generated. Then, such \textit{net-list} can be loaded into a SPICE-like solver where it can be coupled with other circuits. This can be particularly useful when one wants to couple  the \textit{net-list} representing the electrostatic problem  with complex circuit components that are described by dedicated networks provided by the SPICE-like solver but not accessible to the user (e.g., varistor-based
stress control of high-voltage surge arresters). In the following, it is described how to automatically construct a \textit{net-list} equivalent to \eqref{eq:Pq=phi}.

For electrostatic problems, the standard PEEC primary circuit cell \cite{6514777} can be simplified with respect to the one related to full-Maxwell problems.
For the sake of clarity, the very simple problem showed in Fig.~\ref{Fig.patches} can be considered, where two thin and curved conductive domains are shown.  Each one of the two domains is discretized by two patches. Although very simple, this model allow us to  discuss how to automatically construct the  \textit{net-list} for general models. 
It is worth noting that, although the formulation described in this paper has been developed for electrostatic problems only, once the \textit{net-list} equivalent to \eqref{eq:Pq=phi} is constructed and loaded into the SPICE-like solver, the \textit{net-list} can be coupled with other circuits and also frequency or time domain simulations can be performed. Thus, for the sake of generality, in  this section the discussion is carried on by considering a frequency domain problem since the extension is straightforward. 
Fig.~\ref{fig:PEEC_branch} shows the equivalent circuit of the model in Fig.~\ref{Fig.patches}. 
All the  patches that make up a distinct domain are  equipotential, therefore the equivalent circuit has a number of nodes equal to the number of separate domains plus one node for the ground which is at an infinite distance from the devices. The number of patches is indicated with $N$, and in this case $N=4$.

Thus, the electric potential value of  patch $\Gamma_1$ and patch $\Gamma_2$ in  Fig.~\ref{Fig.patches} is given by the potential of node 1 in Fig.~\ref{fig:PEEC_branch}. Analogously, the electric potential value of  patch $\Gamma_3$ and patch $\Gamma_4$ in Fig.~\ref{Fig.patches} is given by the potential of node 2 in Fig.~\ref{fig:PEEC_branch}.

The connections between the patches (circuit nodes) and the ground node are instead modelled with equivalent capacitances and Current Controlled Current Sources (CCCSs). The equivalent capacitances represent the electrostatic interaction between a single patch and the infinity and they are given by 
\begin{equation}
	C_{ii}=\frac{1}{P_{ii}}.
\end{equation}
Instead, the CCCSs represent the electrostatic interactions between different patches. It is worth noting that each CCCS represented in Fig.~\ref{fig:PEEC_branch} should be modelled by $N-1$ CCCSs representing the mutual interaction between two patches (see Fig.~\ref{fig:PEEC_CCCS}) with gain $X$ given by the ratio between the mutual potential coefficient of the two patches and the self potential coefficients of the patches connected to the CCCS, i.e.,
\begin{equation}
	X_{ij}=\frac{P_{ij}}{P_{ii}}.
\end{equation}

  It is worth noting that each  CCCS component at the right part of Fig.~\ref{fig:PEEC_CCCS}  is controlled by the current $i\omega q_i$. At the circuit level, this current can be obtained by summing up the contributions of the current flowing through  the $i$th capacitance and all currents flowing in the CCCSs connected in parallel to the capacitance. Thus, as described in \cite{809841}, to make this current available in the  \textit{net-list}, a Voltage Source (VS) imposing a zero voltage must be added in series to each capacitance. Thus, by taking Fig.~\ref{fig:PEEC_branch} as example, 4 VSs imposing zero voltage are added to the \textit{net-list} and they are connected to the upper terminal of each capacitance and to nodes $n_1$ and $n_2$ as shown in  Fig.~\ref{fig:PEEC_CCCS}.  

Finally, once the PEEC potential matrix is constructed it is possible to automatically generate the equivalent \textit{net-list} by using Algorithm \ref{al:netlist}, where all the required components described above (i.e., capacitances, CCCSs, and zero voltage VSs) are stamped.

\begin{algorithm}[t]
	\caption{Automatic \textit{net-list} generation.} \label{al:netlist}
	\begin{algorithmic}[] 
		\STATE \textbf{Input}: Matrix $\mathbf{P}$ and domain tag for each patch 
		\STATE 
		\STATE Generate void \texttt{netlist.cir} file  \COMMENT{0}
		\STATE 
		\STATE Stamping capacitances and zero VSs in series \COMMENT{1}
		\FOR{i=1,$\cdots$,N} 
			\STATE Find domain tag of the $i$-th patch, i.e., \\d=$domain\_tag$(i)
			\STATE In a new line of \texttt{netlist.cir}, print: \\ \texttt{C\_\#i} \ \texttt{0} \ \texttt{\#i} \ $P^{-1}_{ii}$;
			\STATE In a new line of \texttt{netlist.cir}, print: \\ \texttt{V\_\#i} \ \texttt{\#i} \ \texttt{\#d} \ \texttt{0}; 
		\ENDFOR	
		\STATE
		\STATE Stamping CCCSs \COMMENT{2}
		\FOR{i=1,$\cdots$,N} 
			\FOR{j=1,$\cdots$,N} 
				\STATE In a new line of \texttt{netlist.cir}, print: \\ \texttt{F\_\#i\#j} \ \texttt{0} \ \texttt{\#i} \ \texttt{V\_\#j} \ ${P_{ji}}/{P_{ii}}$;
			\ENDFOR
		\ENDFOR
		\STATE 
		\STATE If required, add other circuit components \COMMENT{3}
		\STATE 
		\STATE 
		\textbf{Output}: \textit{net-list} printed in \texttt{netlist.cir} file
	\end{algorithmic}
\end{algorithm}

From the discussion above, it can be inferred that the generated  \textit{net-list} consist of a large number of components since all the mutual couplings between the patches require a circuit component. As discussed in \cite{Safavi2012}, this can become an issue when a large number of unknowns is required since standard SPICE-like solvers may be inefficient when large systems of equations with dense matrix blocks are generated. However, it is worth noting that the high-order curved geometry description allows for keeping small the number of required patches, thus mitigating such issue. 

\subsection{Convergence} \label{sec:convergence}
For sufficiently smooth geometries $\Gamma$ and solutions $\varphi$ of \eqref{eq:poisson} the expected convergence order of the approximation $\rho_h$ is $\mathcal{O}(h^{3/2 + p})$. This is a consequence of the convergence theory for boundary element methods \cite[Cor.~4.1.34]{Sauter_2011aa}.
The error of the potential evaluated in \eqref{eq:phie} by inserting $\rho_h$ converges even faster, i.e. with $\mathcal{O}(h^{3 + 2p})$ due to the Aubin-Nitsche trick \cite[Thm.~4.2.14]{Sauter_2011aa}.
This trick holds more generally for any linear functional of the approximation $\rho_h$ that can be evaluated with sufficient accuracy. 
Since we obtain the capacities, see \eqref{eq:Pcoeff}, in the isogeometric approach by surface integration based on the exact geometry, we also profit here from the increased convergence rate.

\section{Numerical Results} \label{sec:num_res}

In this section, simulation results obtained from  numerical experiments are shown.
Two test cases are considered. The first is an academic example consisting of a spherical capacitor, i.e., two concentric spheres. While being simple, this test case allow us to show the benefits of the proposed spline-based PEEC approach due to its strong curvature. Moreover, this problem has a closed-form solution and therefore we can test and compare the accuracy of the numerical methods.

The second test case instead consists on the realistic model of a surge arrester. In this model, thanks to the  circuit interpretation provided by the PEEC scheme, the surge arrester model is excited by using a  lumped voltage source.

The proposed PEEC method has been implemented within the Bembel, The fast isogeometric boundary element {C++} library~\cite{Dolz_2020ac}. The implementation provides an interface to the Eigen template library for linear algebra operations \cite{eigen3}. Furthermore, the matrix assembly is openMP parallelized~\cite{OpenMP-Architecture-Review-Board_2008aa}. For comparison we also use the Matlab implementation of PEEC introduced in \cite{8764572} based on low-order surface mesh (i.e., triangular elements)  and zero-order shape functions. All code is executed on a cluster with Intel\textregistered \ Xeon\textregistered \ Platinum 8160 CPU @ 2.10GHz.

\begin{figure}[t!]
	\centering
	\begin{tikzpicture}
		\pgfplotsset{
		   every axis/.append style={
				font=\fontsize{10}{10}\sffamily},
			every non boxed x axis/.append style={
				x axis line style={->}
			},
			every non boxed y axis/.append style={
				y axis line style={->}
			},
			every non boxed z axis/.append style={
				z axis line style={->}
			}
		}
			\begin{loglogaxis}[
				height = 5.5cm,
				width = 0.95\linewidth,
				xmin = 5,
				xmax = 1e4,
				ymax = 1e-1,
				ymin = 1e-8,
				ytick distance=100,
				xlabel=Number of degrees of freedom,
				ylabel=Relative error in $C$,
				ytick = {1e-1,1e-3,1e-5,1e-7},
				yticklabels = { 10\textsuperscript{-1},
								10\textsuperscript{-3},
								10\textsuperscript{-5},
								10\textsuperscript{-7}},
				xtick = {1e1,1e2,1e3,1e4},
				xticklabels = { 10\textsuperscript{1},
								10\textsuperscript{2},
								10\textsuperscript{3},
								10\textsuperscript{4}},
				legend style={at={(0.0,0.0)},anchor=south west},
				]
				\addplot[color=red,mark=*] coordinates {
					(12,		0.0328329)
					(48,		0.000891369)
					(192,		2.47151e-05)
					(3072,	1.10885e-08)
				};
				\addlegendentry{Splines}
				
				\addplot[color=blue,mark=*] coordinates {
					(104,0.068)
					(416,0.0195)
					(1664,0.0051)
					(6656,0.0013)
					
				};
				\addlegendentry{Triangles}

				\addplot+[black,dashed,no marks,samples at={5,10000}] {10^(3-3*log10(x))};
				\addplot+[black,dashed,no marks,samples at={5,10000}] {0.3*10^(1-1*log10(x))};
				\node at (axis cs:500,8e-7) {$\mathcal{O}(h^3)$};
				\node at (axis cs:5000,1e-4) {$\mathcal{O}(h)$};
			\end{loglogaxis}
		\end{tikzpicture}
	\caption{Convergence analysis of the relative error of the numerically computed capacity plotted against the utilized degrees of freedom. The new method is denoted by `Splines' and a conventional reference implementation by `Triangles'.}
	\label{fig:numerical}
\end{figure}
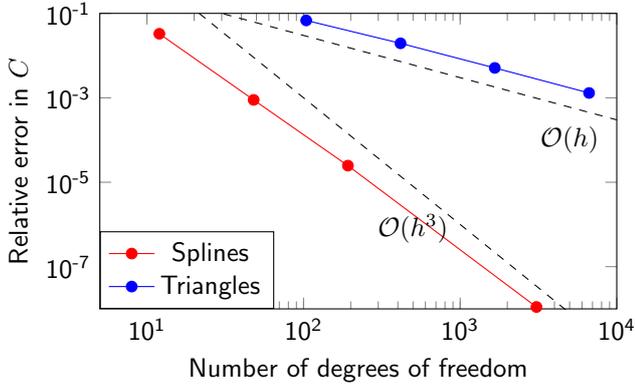

% Define block styles
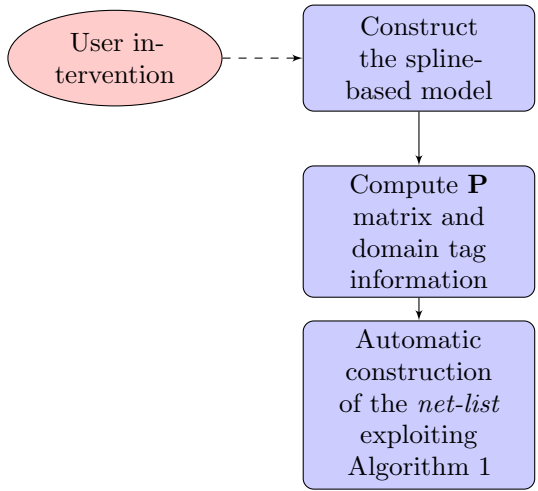
\begin{figure}
    \centering
\tikzstyle{decision} = [diamond, draw, fill=blue!20, 
    text width=4.5em, text badly centered, node distance=3cm, inner sep=0pt]
\tikzstyle{block} = [rectangle, draw, fill=blue!20, 
    text width=8em, text centered, rounded corners, minimum height=4em]
\tikzstyle{line} = [draw, -latex']
\tikzstyle{cloud} = [draw, ellipse,fill=red!20, node distance=4cm, text width=5em, text centered,
    minimum height=2em]
\begin{tikzpicture}[node distance = 2.3cm, auto]
    \node [block] (model) {Construct the spline-based model};
    \node [cloud, left of=model] (user) {User intervention};
    \node [block, below of=model] (matrices) {Compute $\mathbf{P}$ matrix and domain tag information};
    \node [block, below of=matrices] (netlist) {Automatic construction of the \textit{net-list} exploiting Algorithm~\ref{al:netlist} };
    \path [line] (model) -- (matrices);
    \path [line] (matrices) -- (netlist);
    \path [line,dashed] (user) -- (model);
\end{tikzpicture}
\caption{Work-flow for the automatic \textit{net-list} generation.} \label{Fig.flow}
\end{figure}

\subsection{Academic Example: Two Concentric Spheres}
The method has been applied to the electrostatic configuration of two concentric perfect electric conducting spherical shells of radii $r_{\mathrm{in}}=\SI{0.1}{\meter}$ and $r_{\mathrm{out}}=\SI{0.2}{\meter}$.
The model shown in Fig.~\ref{fig:capa} consists of 12 NURBS patches. Without further refinement, this leads to 78 partial capacitances with total capacitance of $C=\SI{21.5}{\pico\farad}$. The reference value is the closed-form solution \cite{Jackson_1995aa} 
$$C_{\mathrm{ana}} = \frac{4\pi\varepsilon_0}{\left(1/r_{\mathrm{in}} - {1}/{r_{\mathrm{out}}}\right)}\approx\SI{22.25}{\pico\farad}.$$
The slopes for the isogeometric method shown in Fig.~\ref{fig:numerical} obtained by mesh refinement meet the expected optimal convergence order of $\mathcal{O}(h^3)$ for lowest-order basis functions, see Section~\ref{sec:convergence}. On the other hand, Fig.~\ref{fig:numerical} also shows the results obtained from the standard PEEC method based on low-order surface mesh (i.e., triangular elements) and zero-order shape functions. Here, the convergence is limited to first order since we do not evaluate the surface integral with sufficient accuracy and the Aubin-Nitsche trick is not applicable. This is well-known in the context of boundary element methods, see for example \cite[Tbl.~8.2]{Sauter_2011aa} for an overview of the necessary orders of curvature approximations. The two curves in Fig.~\ref{fig:numerical} confirm the theoretical prediction that the convergence order of the proposed spline based PEEC method is much higher than (non-curved) triangle-based PEEC approaches. 

The results  above and in Fig.~\ref{fig:numerical} were obtained by solving \eqref{eq:Pq=phi} with the LU decomposition implemented in Eigen \cite{eigen3}. However, for the sake of completeness, the automatic construction of  the \textit{net-list}
as discussed in Section \ref{sec:Stamping} and schematically reported in Fig.~\ref{Fig.flow} is also performed. The generated \textit{net-list} is loaded into the SPICE-Solver LTspice and solved by executing the circuit simulator. The obtained results are, as expected, identical to the one obtained from LU decomposition up to machine precision. A part of the generated  \texttt{netlist.cir} file is shown in Fig.~\ref{Fig.netlist} for illustration; the code is available in \cite{link-to-the-repo}.

\begin{figure} [t!]
	\centering
	\tiny\begin{verbatim}
* my netlist

C 1 0 1 5.0964e-11; capacitance of patch #1 w.r.t. infinity
C 2 0 2 5.0964e-11; capacitance of patch #2 w.r.t. infinity
C 3 0 3 5.0964e-11; capacitance of patch #3 w.r.t. infinity
C 4 0 4 5.0964e-11; capacitance of patch #4 w.r.t. infinity
C 5 0 5 5.0964e-11; capacitance of patch #5 w.r.t. infinity
C 6 0 6 5.0964e-11; capacitance of patch #6 w.r.t. infinity
C 7 0 7 1.0193e-10; capacitance of patch #7 w.r.t. infinity
C 8 0 8 1.0193e-10; capacitance of patch #8 w.r.t. infinity
C 9 0 9 1.0193e-10; capacitance of patch #9 w.r.t. infinity
C 10 0 10 1.0193e-10; capacitance of patch #10 w.r.t. infinity
C 11 0 11 1.0193e-10; capacitance of patch #11 w.r.t. infinity
C 12 0 12 1.0193e-10; capacitance of patch #12 w.r.t. infinity
V 1 13 1 O; 0-voltage source to measure the displacement current in patch #1
V 2 13 2 0; 0-voltage source to measure the displacement current in patch #2
V 3 13 3 O; 0-voltage source to measure the displacement current in patch #3
V 4 13 4 O; 0-voltage source to measure the displacement current in patch #4
V 5 13 5 O; 0-voltage source to measure the displacement current in patch #5
V 6 13 6 O; 0-voltage source to measure the displacement current in patch #6
V 7 14 7 O; 0-voltage source to measure the displacement current in patch #7
V 8 14 8 O; 0-voltage source to measure the displacement current in patch #8
V 9 14 9 O; 0-voltage source to measure the displacement current in patch #9
V 10 14 10 0; 0-voltage source to measure the displacement current in patch #10
V 11 14 11 O; 0-voltage source to measure the displacement current in patch #11
V 12 14 12 O; 0-voltage source to measure the displacement current in patch #12
F 1 2 0 1 V 2 0.39693; current control current source #1_2
F 1 3 0 1 V 3 0.39693; current control current source #1_3
F 1 4 0 1 V 4 0.39693; current control current source #1_4
F 1 5 0 1 V 5 0.39693; current control current source #1_5
F 1 6 0 1 V 6 0.26034; current control current source #1_6
F 1 7 0 1 V 7 0.34106; current control current source #1_7
F 1 8 0 1 V 8 0.22821; current control current source #1_8
F 1 9 0 1 V 9 0.22821; current control current source #1_9
F 1 10 0 1 V 10 0.22821; current control current source #1_10
F 1 11 0 1 V 11 0.22821; current control current source #1_11
F 1 12 0 1 V 12 0.17144; current control current source #1_12
F 2 1 0 2 V 1 0.39693; current control current source #2_1
F 2 3 0 2 V 3 0.39693; current control current source #2_3
F 2 4 0 2 V 4 0.26034; current control current source #2_4
F 2 5 0 2 V 5 0.39693; current control current source #2_5
	\end{verbatim}
	\vspace*{-1.5em}
	\caption{First 41 lines of the generated \texttt{netlist.cir} file for the concentric sphere shells test case.}
	\label{Fig.netlist}
\end{figure}

\subsection{Surge Arrester}

\begin{figure}  [t!]
	\centering
	\includegraphics[trim=0mm 0mm 0mm 0mm, clip,width=1\columnwidth]{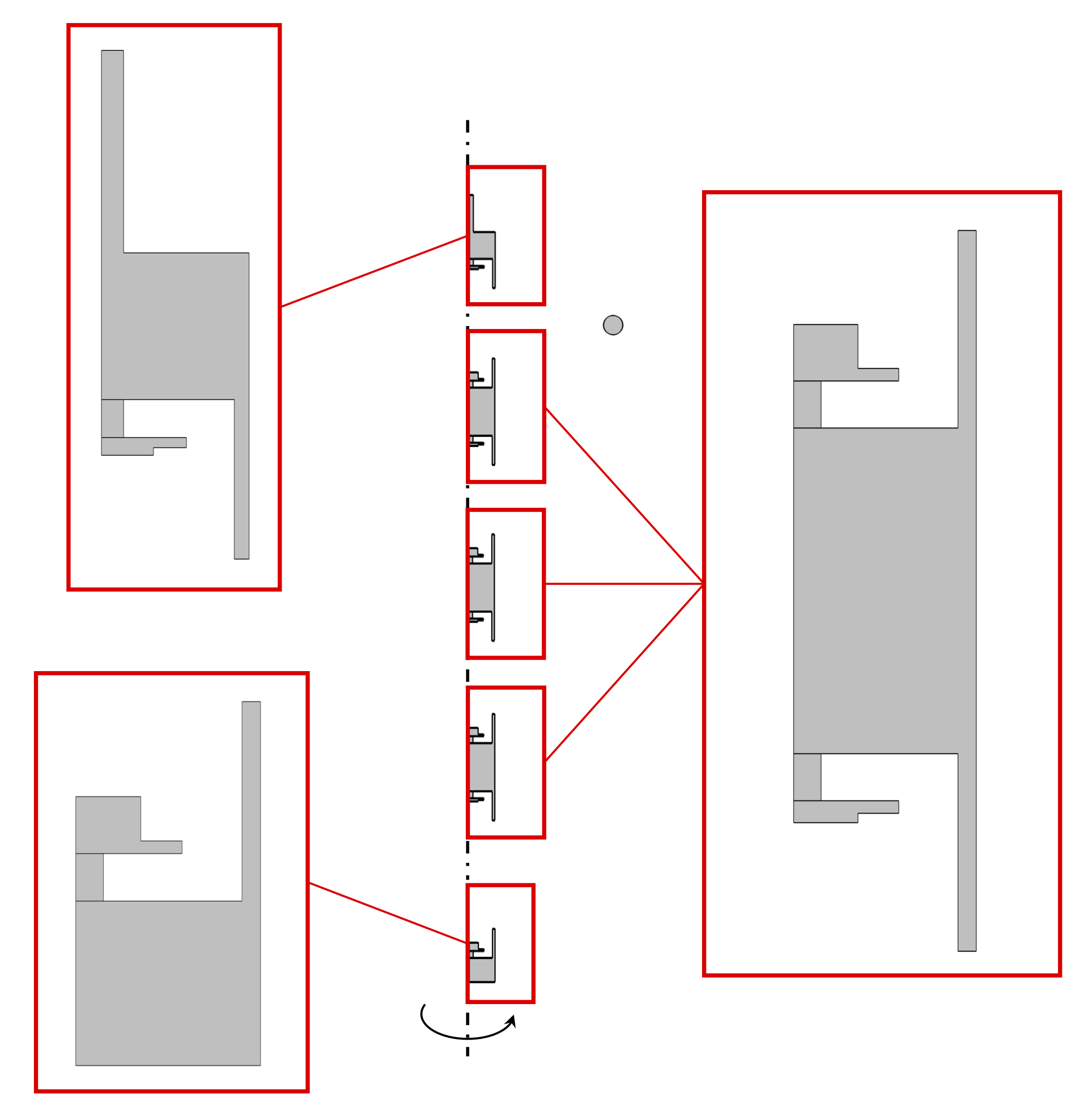}
	\caption{Axisymmetric representation of the surge arrester model derived from \cite{8630588}, with grading ring from \cite[Tbl.~4.1]{Giessel_2018aa}, not to scale.}
	\label{fig:geo_high_voltage}
\end{figure}

In this section, the model of a realistic surge arrester is considered. The geometry and the dimensions of the model are shown in Fig.~\ref{fig:geo_high_voltage} and they are derived from \cite{8630588}.
The model consists on five metallic domains placed one over the other plus a conductive ring. 
 A voltage source applying \SI{1}{\volt} is placed between the top and the bottom metallic domain. Moreover, the ring and the top metallic domain are connected with a circuit short thus they are at the same potential. 
 
\begin{figure} [t!]
	\centering
	\includegraphics[width=0.6\columnwidth,trim=0 0 0 0, clip]{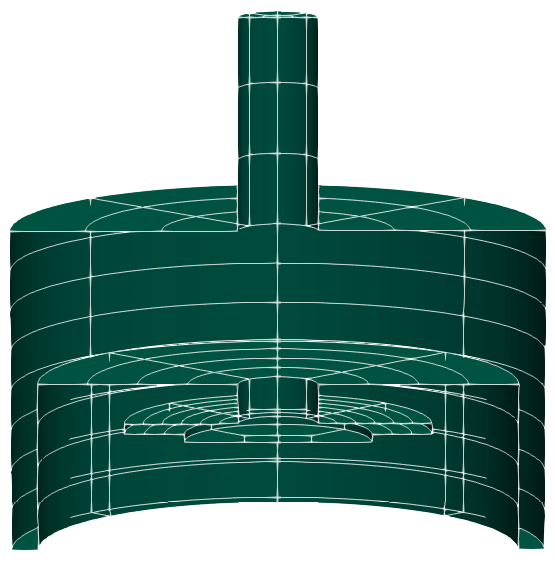}
	\caption{Exemplary representation of top domain of the surge arrester in the NURBS parametrization.}
	\label{fig:mesh_nurbs_top}
\end{figure}

\begin{figure}[t!]
	\centering
	\includegraphics[width=\columnwidth,trim=0 0 0 0, clip]{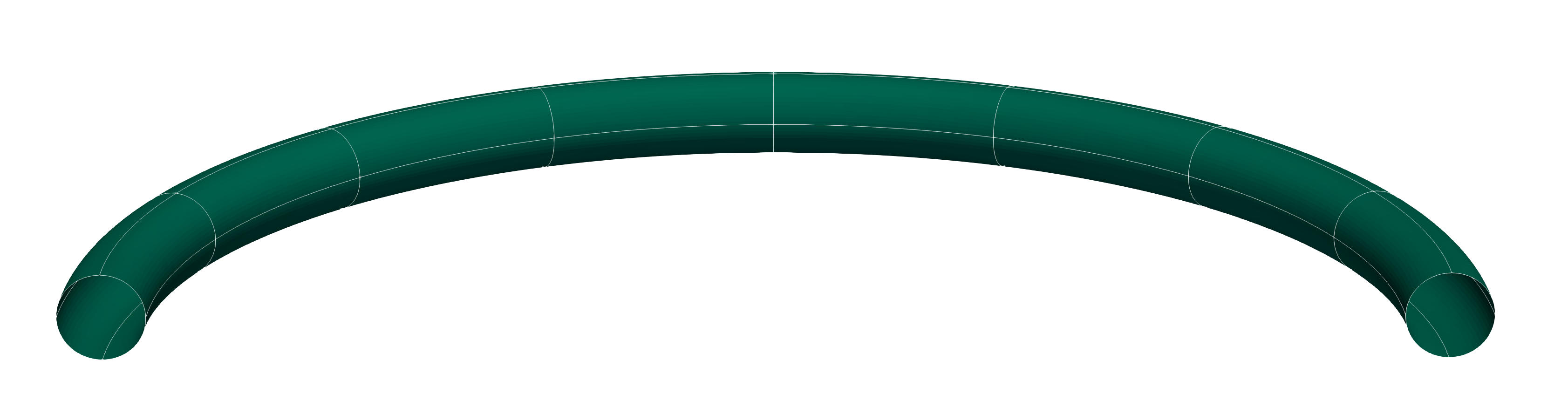}
	\caption{Exemplary representation of the conductive ring of the surge arrester in the NURBS parametrization.}
	\label{fig:mesh_nurbs_ring}
\end{figure}

Using the spline-based PEEC method, the model consists of 1912 patches, resulting in 30592 elements with twofold uniform refinement.
Fig.~\ref{fig:mesh_nurbs_ring} and  Fig.~\ref{fig:mesh_nurbs_top} show a detail of the conductive ring and top domain  spline based PEEC model, respectively.

By using the standard PEEC method with triangular elements, the model consists on 30113 degrees of freedoms. The capacitance value obtained by using a classical Finite Element Method approach implemented in the commercial software COMSOL\textregistered \ is considered as external reference. The (axisymmetric) FEM model consists of 883297 degrees of freedoms and cubic elements.
 The value of equivalent capacitance of the structure obtained from the (axisymmetric) FEM model and the spline and triangular based PEEC methods are in very good agreement, i.e., \SI{9.2935}{\pico\farad}, \SI{9.2942}{\pico\farad}, and \SI{9.2715}{\pico\farad}, respectively.

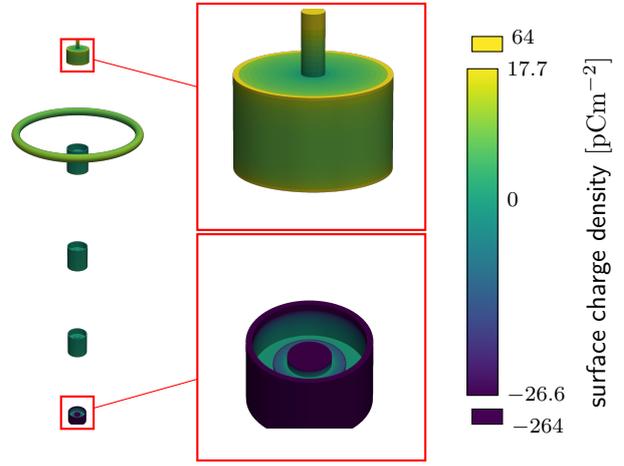
\begin{figure}  [t!]
	\begin{tikzpicture}[spy using outlines={rectangle, magnification=7, connect spies}]
		\pgfplotsset{
			every axis/.append style={font=\fontsize{10}{10}\sffamily},
			every non boxed x axis/.append style={x axis line style={->}},
			every non boxed y axis/.append style={y axis line style={->}},
			every non boxed z axis/.append style={z axis line style={->}}
		}
		\begin{axis}[
		width=0.43\textwidth,
		colorbar,
		colorbar style={
			ytick={-26.6, 0, 17.7},
			ticklabel style={font=\footnotesize},
			ylabel=surface charge density $[\si{\pico\coulomb\metre^{-2}}]$,
			at={(1,0.5)},
			anchor=west,
			height=0.85*\pgfkeysvalueof{/pgfplots/parent axis height},
		},
		colorbar/width=4mm,
		colormap name=viridis,
		point meta min=-26.6,
		point meta max=17.7,
		axis line style={draw=none},
		tick style={draw=none},
		xtick=\empty,
		ytick=\empty,
		ymax=5,
		ymin=-5,
		xmax=5,
		xmin=-1,
		]
			\node at (0,0) {\includegraphics[width=1.9cm,trim={0 2000 0 750},clip]{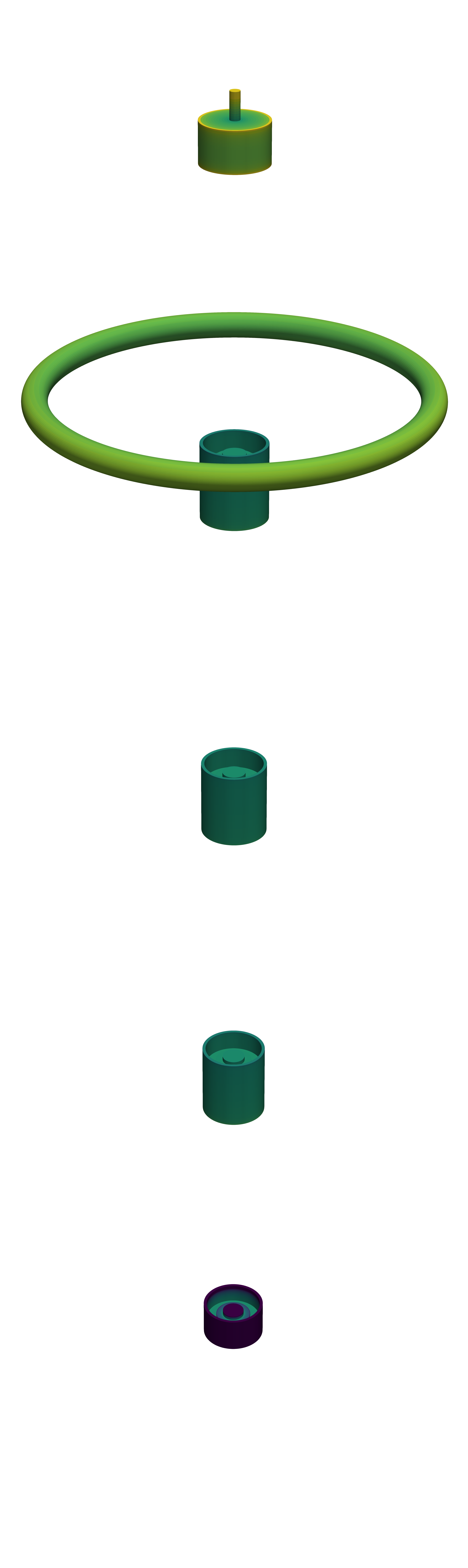}};
			\coordinate (spypoint1) at (0,4.6);
			\coordinate (magnifyglass1) at (3,3);
			\coordinate (spypoint2) at (0,-4.7);
			\coordinate (magnifyglass2) at (3,-3); 
			\spy [red, size=3cm, every spy on node/.append style={thick}] on (spypoint1) in node[fill=white] at (magnifyglass1);
			\spy [red, size=3cm, every spy on node/.append style={thick}] on (spypoint2) in node[fill=white] at (magnifyglass2);
		\end{axis}
		\fill[color of colormap=1000 of viridis, draw=black] (6.22,4.94) rectangle ++(4mm,2mm)node[black,font=\footnotesize,anchor=west] {$64$};
		\fill[color of colormap=0 of viridis, draw=black] (6.22,0) rectangle ++(4mm,2mm)node[black,font=\footnotesize,anchor=north west] {$-264$};
	\end{tikzpicture}
	\caption{Visualization of the computed surface charge density distribution obtained from the spline-based PEEC method.}
	\label{fig:solution}
\end{figure}

\begin{figure}
	\centering
	\begin{tikzpicture}
		\pgfplotsset{
			every axis/.append style={font=\fontsize{10}{10}\sffamily},
			every non boxed x axis/.append style={x axis line style={->}},
			every non boxed y axis/.append style={y axis line style={->}},
			every non boxed z axis/.append style={z axis line style={->}}
		}
		\pgfplotsset{
			colormap={parula}{
				rgb255=(53,42,135)
				rgb255=(15,92,221)
				rgb255=(18,125,216)
				rgb255=(7,156,207)
				rgb255=(21,177,180)
				rgb255=(89,189,140)
				rgb255=(165,190,107)
				rgb255=(225,185,82)
				rgb255=(252,206,46)
				rgb255=(249,251,14)}}
		\begin{axis}[
			width=0.43\textwidth,
			colorbar,
			colorbar style={
				ytick={-250,-200,...,0},
				ticklabel style={font=\footnotesize},
				ylabel=surface charge density $[\si{\pico\coulomb\metre^{-2}}]$,
			},
			colormap name=parula,
			point meta min=-260,
			point meta max=0,
			axis line style={draw=none},
			tick style={draw=none},
			xtick=\empty,
			ytick=\empty,
			ymax=4,
			ymin=-4,
			xmax=2,
			xmin=-2,
		]
			\node at (0,0) {\includegraphics[width=5cm,trim={220 30 90 30},clip]{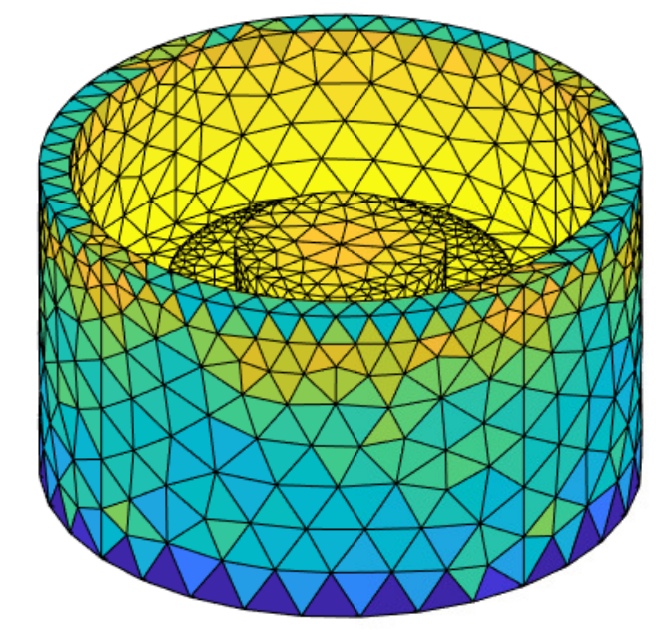} };
		\end{axis}
	\end{tikzpicture}
	\caption{Visualization of the computed surface charge density distribution obtained from the  standard PEEC method.}
	\label{fig:charge_sPEEC}
\end{figure}

Finally, Fig.~\ref{fig:solution} and Fig.~\ref{fig:charge_sPEEC} show the surface charge density distribution on the structure obtained from from the spline and triangular based PEEC methods, respectively. As can be seen, thanks to the perfect representation of the curved geometry by means of splines, the charge density distribution obtained from the spline based PEEC method is smoother than the one obtained from the triangular based PEEC approach.

\section{Conclusions}
This paper proposes a Partial Element Equivalent Circuit (PEEC)  method based on the function spaces from Isogeometric Analysis (IgA). It is applied to electrostatics.
Thanks to the use of an integral equation formulation, the adoption of spline-based geometry concepts from IgA and piece-wise constant basis functions, it is shown how a fully-coupled capacitance circuit can be extracted bypassing the usual costly meshing step.
This saves time and manual efforts of a potential user.
It is demonstrated that a realistic problem with curved geometry can be solved with adequate accuracy. Moreover, the proposed IgA-PEEC
method converges for curved geometries up to three
times faster than a conventional PEEC implementation.

\end{document}